\documentstyle[aps,preprint]{revtex}

\draft

\begin{document}
\title{Spin-Flip Scattering Effect on the Current-Induced Spin Torque in
Ferromagnet-Insulator-Ferromagnet Tunnel Junctions}
\author{Zhen-Gang Zhu, Gang Su$^{\ast }$, Biao Jin, and Qing-Rong Zheng}
\address{Department of Physics, The Graduate School of the Chinese Academy of\\
Sciences, P.O. Box 3908, Beijing 100039, China}
\maketitle

\begin{abstract}
We have investigated the current-induced spin transfer torque of a
ferromagnet-insulator-ferromagnet tunnel junction by taking the spin-flip
scatterings into account. It is found that the spin-flip scattering can
induce an additional spin torque, enhancing the maximum of the spin torque
and giving rise to an angular shift compared to the case when the spin-flip
scatterings are neglected. The effects of the molecular fields of the left
and right ferromagnets on the spin torque are also studied. It is found that 
$\tau ^{Rx}/I_{e}$ ($\tau ^{Rx}$ is the spin-transfer torque acting on the
right ferromagnet and $I_{e}$ is the tunneling elcetrical current) does vary
with the molecular fields. At two certain angles, $\tau ^{Rx}/I_{e}$ is
independent of the molecular field of the right ferromagnet, resulting in
two crossing points in the curve of $\tau ^{Rx}/I_{e}$ versus the relevant
orientation for different molecular fields.
\end{abstract}


The spin-polarized transport in multilayer structures exhibits new effects
such as the giant magnetoresistance \cite{wolf} (GMR), the spin transfer
effect \cite{slonczewski}, and so on. How to use the spin degree of freedom
of electrons in ferromagnetic materials to construct new devices is at
present a focus in the field of spintronics. Spin-polarized electrons
flowing from one ferromagnetic layer into another layer in which the
molecular field deviates by an angle may transfer the angular momentum to
the local angular momentum of the ferromagnetic layer, thereby exerting a
torque on the magnetic moments (see e.g. Refs. \cite
{slonczewski,s2,tsoi,sun,myers,katine,fnf,waintal}). This phenomenon is
usually called the spin transfer effect\cite{slonczewski}. The torques in
the plane spanned by ${\bf s}_{1}$ and ${\bf s}_{2}$, where ${\bf s}_{1}$
and ${\bf s}_{2}$ are spin moments in the left and right ferromagnets, are
normally called the dynamic nonequilibrium spin torques\cite{waintal}. Spin
transfer motion of ${\bf s}_{1}$ and ${\bf s}_{2}$ within their spanned
plane is different from the spin precession like $\partial {\bf s}%
_{1}/\partial t=\hbar J{\bf s}_{1}\times {\bf s}_{2}$ out of the spanned
plane which describes the conventional exchange coupling \cite
{waintal2,erickson}. Therefore, the spin transfer effect causes new physical
phenomena in magnetic multilayer structures. When the current is large
enough, it could switch the magnetic states of the local angular momentum.
Such a current-induced change of the magnetic state has been observed in
several experiments (see e.g. Refs.\cite{tsoi,sun,myers,katine}). As a
result, the spin transfer effect may provide a mechanism for a
current-controlled magnetic memory element. To deal with the spin transfer
effect, it is useful to introduce the concepts such as the spin current and
the spin torque to describe the coupling between the conduction electrons
and the magnetic moments of ferromagnetic materials. These concepts are
first proposed by Slonczewski \cite{s2} based on a quantum-mechanical model
for ferromagnet-insulator-ferromagnet (FM-I-FM) junctions. Then, the
concepts are extended to the structures such as ferromagnet-normal
metal-ferromagnet (FM-NM-FM) junctions \cite{fnf,waintal},
ferromagnet-superconductor-ferromagnet junctions \cite{fsf}, and trilayer
FM-NM-FM which contacts an normal metal lead or a superconductor lead\cite
{waintal1,waintal2}, etc., showing that the investigation on spin torques in
magnetic junctions has been receiving much attention.

Many works concerning the spin torque are presented for FM-NM-FM structures
so far. The work for FM-I-FM structures is still sparse. In particular, when
electrons tunnel through the insulator barrier, the spin-flip scattering may
occur\cite{moodera,vedyayev}. The spin-flip electrons feel a different
torque with respect to those non-flip electrons, and could exert an
additional torque to the ferromagnet. Consequently, this additional torque
induced by the spin-flip electrons may play a role as the dynamic spin
transfer torque. In this paper, we shall use the nonequilibrium Green
function technique\cite{haug} to investigate the spin-flip scattering effect
on the current-induced spin transfer torque in FM-I-FM tunnel junctions.

The system under interest is composed of two ferromagnets which are
stretched to infinite separated by a thin insulator, as illustrated in
Fig.1. The molecular field in the left ferromagnet is assumed to align along
the $z$ axis which is in the junction plane, while the orientation of the
molecular field in the right ferromagnet, along the $z^{\prime }$ axis,
deviates the $z$ axis by an angle $\theta $. The electrons flow along the $x$
axis which is perpendicular to the junction plane. The Hamiltonian of the
system is\ \ \ \ \ \ \ \ \ \ \ \ \ \ \ \ \ \ \ \ \ \ \ \ \ \ \ \ \ \ \ \ \ \
\ \ \ \ \ \ \ \ \ \ \ \ \ \ \ \ \ \ \ \ \ \ \ \ \ \ \ \ \ \ \ \ \ \ \ \ \ \
\ \ \ \ \ \ \ \ \ \ \ \ \ \ \ \ \ \ \ \ \ \ \ \ \ \ \ \ \ \ \ \ \ \ \ \ \ \
\ \ \ \ \ \ \ \ \ \ \ \ \ \ \ \ \ \ \ \ \ \ \ \ \ \ \ \ \ \ \ 

\begin{equation}
H=H_{L}+H_{R}+H_{T},  \label{Thamiltonian}
\end{equation}
with 
\begin{eqnarray}
H_{L} &=&\sum_{k\sigma }\varepsilon _{k\sigma }^{L}a_{k\sigma }^{\dagger
}a_{k\sigma },  \label{respectH} \\
H_{R} &=&\sum_{q\sigma }[(\varepsilon _{R}({\bf q)-\sigma }M_{2}\cos \theta
)c_{q\sigma }^{\dagger }c_{q\sigma }-M_{2}\sin \theta c_{q\sigma }^{\dagger
}c_{q\overline{\sigma }}],  \nonumber \\
H_{T} &=&\sum_{kq\sigma \sigma ^{\prime }}[T_{kq}^{\sigma \sigma ^{\prime
}}a_{k\sigma }^{\dagger }c_{q\sigma ^{\prime }}+T_{kq}^{\sigma \sigma
^{\prime }}{}^{\ast }c_{q\sigma ^{\prime }}^{\dagger }a_{k\sigma }], 
\nonumber
\end{eqnarray}
where $a_{k\sigma }$ and $c_{k\sigma }$ are annihilation operators of
electrons with momentum $k$ and spin $\sigma $ $(=\pm 1)$ in the left and
right ferromagnets, respectively, $\varepsilon _{k\sigma }^{L}=\varepsilon
_{L}({\bf k)-}eV{\bf -\sigma }M_{1},$ $M_{1}=\frac{g\mu _{B}h_{L}}{2},$ $%
M_{2}=\frac{g\mu _{B}h_{R}}{2},$ $g$ is the Land\'{e} factor, $\mu _{B}$ is
the Bohr magneton, $h_{L(R)}$ is the molecular field of the left (right)
ferromagnet, $\varepsilon _{L(R)}({\bf k)}$ is the single-particle
dispersion of the left (right) FM electrode, $V$ is the applied bias
voltage, $T_{kq}^{\sigma \sigma ^{\prime }}$ denotes the spin and momentum
dependent tunneling amplitude through the insulating barrier. Note that the
spin-flip scattering is included in $H_{T}$ when $\sigma ^{\prime }=\bar{%
\sigma}=-\sigma $. It is this term that violates the spin conservation in
the tunneling process.

With the system defined above, let us now consider the spin torques exerting
on the magnetic moments in the {\it right} FM electrode of this magnetic
tunnel junction. The spin torques, namely the time evolution rate of the
total spin of the left or the right ferromagnet, can be obtained by $\frac{%
\partial }{\partial t}\langle {\bf s}_{1,2}(t)\rangle =\frac{i}{\hbar }%
\langle \lbrack H,$ ${\bf s}_{1,2}(t)]\rangle .$ In Refs.\cite
{waintal,waintal1}, the spin torques are defined by considering the momentum
conservation $\partial {\bf s}_{1}/\partial t={\bf I(-\infty )-I(}0)$\cite
{slonczewski}, where ${\bf I}$ is the spin current (whose definition can be
found in Refs.\cite{waintal2,fnf}). Because there are the spin-dependent
scatterings caused by the local exchange field inside the ferromagnets, the
spin current is no longer conserved inside the ferromagnets. Whereas the
total spin is conserved, the lost spin current is transferred to the local
magnetic moments, thereby giving rise to a torque exerting on the local
magnetic moments of the ferromagnets. So the nonconservation of the
nonequilibrium spin current leads to a current-induced nonequilibrium torque 
\cite{waintal2}. We may consider this issue in another way, i.e. to
investigate the evolution rate of the total spin of the ferromagnets. In
doing so, one must be cautious to identify the spin-torques implied from the
total Hamiltonian. In this model one may see that the right ferromagnet
gains two types of torques: one is the equilibrium torque caused by the
spin-dependent potential ( i.e. the magnetic exchange interaction), and the
other is from the electrons tunneling through the insulating barrier from
the left side. The latter can be obtained from the tunneling term $H_{T}$,
which is nothing but the current-induced spin transfer torque in the plane
spanned by ${\bf h}_{L}$ and ${\bf h}_{R}$. When the applied bias is absent,
the left and right ferromagnets will only undergo the spin torque caused by
the spin-dependent potential, and the current-induced torques would not
appear. Therefore, we can define the current-induced spin transfer torque as 
${\bf \tau }=\frac{i}{\hbar }\langle \lbrack H_{T},$ ${\bf s}_{2}(t)]\rangle 
$. Note that a similar expression in a current matrix form is introduced in
Ref.\cite{fsf}. The equilibrium torques caused by a spin-dependent potential
will not be considered here. The total spin of the right ferromagnet is 
\begin{equation}
{\bf s}_{2}(t)=\frac{\hbar }{2}%
\mathrel{\mathop{\sum }\limits_{k\mu \nu }}%
c_{k\mu }^{\dagger }c_{k\nu }({\bf R}^{-1}\chi _{\mu })^{\dagger }\stackrel{%
\wedge }{{\bf \sigma }}({\bf R}^{-1}\chi _{\nu }),  \label{totalsp}
\end{equation}
where ${\bf R}=\left( 
\begin{array}{cc}
\cos \frac{\theta }{2} & -\sin \frac{\theta }{2} \\ 
\sin \frac{\theta }{2} & \cos \frac{\theta }{2}
\end{array}
\right) $, $\stackrel{\wedge }{{\bf \sigma }}$ is Pauli matrices and $\chi
_{\mu (\nu )}$ is spin states. Note that Eq. (\ref{totalsp}) is written in
the $xyz$ coordinate frame while the spins ${\bf s}_{2}$ are quantized in
the $x^{\prime }y^{\prime }z^{\prime }$ frame. From ${\bf \dot{s}}_{1,2}\sim
I_{e}\widehat{s}_{1,2}\times (\widehat{s}_{1}\times \widehat{s}_{2})$ with $%
I_{e}$ the electrical current, we can judge that the direction of the spin
transfer torque is along $x^{\prime }$ direction in the $x^{\prime
}y^{\prime }z^{\prime }$ coordinate frame. We can further write the spin
transfer torque in Eq.(\ref{totalsp}) as ${\bf s}_{2}(t)=\frac{\hbar }{2}%
\mathrel{\mathop{\sum }\limits_{k\sigma }}%
(c_{k\sigma }^{\dagger }c_{k\overline{\sigma }}\cos \theta -\sigma
c_{k\sigma }^{\dagger }c_{k\sigma }\sin \theta )=s_{2x^{\prime }0}\cos
\theta -s_{2z^{\prime }0}\sin \theta ,$ where $s_{2x^{\prime }0}$ and $%
s_{2z^{\prime }0}$ are $x^{\prime }$- and $z^{\prime }$-components of the
total spins in the $x^{\prime }y^{\prime }z^{\prime }$ coordinate frame in
which the spins ${\bf s}_{2}$ are quantized. Hence, the current-induced spin
transfer torque can be obtained:

\begin{equation}
\tau ^{Rx^{\prime }}=-\cos \theta 
\mathop{\rm Re}%
\mathrel{\mathop{\sum }\limits_{kq}}%
\int \frac{d\varepsilon }{2\pi }Tr_{\sigma }[{\bf G}_{kq}^{<}(\varepsilon )%
\stackrel{\wedge }{\sigma }_{1}{\bf T}^{\dagger }]+\sin \theta 
\mathop{\rm Re}%
\mathrel{\mathop{\sum }\limits_{kq}}%
\int \frac{d\varepsilon }{2\pi }Tr_{\sigma }[{\bf G}_{kq}^{<}(\varepsilon )%
\stackrel{\wedge }{\sigma }_{3}{\bf T}^{\dagger }],  \label{torquex}
\end{equation}
where $\stackrel{\wedge }{\sigma }_{1}=\left( 
\begin{array}{cc}
0 & 1 \\ 
1 & 0
\end{array}
\right) ,$ $\stackrel{\wedge }{\sigma }_{3}=\left( 
\begin{array}{cc}
1 & 0 \\ 
0 & -1
\end{array}
\right) $ are Pauli matrices, ${\bf T}=\left( 
\begin{array}{cc}
T_{1} & T_{2} \\ 
T_{3} & T_{4}
\end{array}
\right) $ with the elements $T_{i}$ ($i=1,...,4)$ being the tunneling
amplitudes which are for simplicity assumed to be independent of $k$ and $q$
(namely, $T^{\uparrow \uparrow }$, $T^{\uparrow \downarrow }$, $%
T^{\downarrow \uparrow }$, $T^{\downarrow \downarrow }$ respectively), $%
Tr_{\sigma }$ stands for the trace of the matrix taking over the spin space,
and ${\bf G}_{kq}^{<}(\varepsilon )$ is the lesser Green function in spin
space defined as

\begin{equation}
{\bf G}_{kq}^{<}(\varepsilon )=\left( 
\begin{array}{cc}
G_{kq}^{\uparrow \uparrow ,<}(\varepsilon ) & G_{kq}^{\downarrow \uparrow
,<}(\varepsilon ) \\ 
G_{kq}^{\uparrow \downarrow ,<}(\varepsilon ) & G_{kq}^{\downarrow
\downarrow ,<}(\varepsilon )
\end{array}
\right) ,  \label{lesserGF}
\end{equation}
with $G_{kq}^{\sigma \sigma ^{\prime },<}(\varepsilon )=\int
dte^{i\varepsilon (t-t^{\prime })}G_{kq}^{\sigma \sigma ^{\prime
},<}(t-t^{\prime }),$ and $G_{kq}^{\sigma \sigma ^{\prime },<}(t-t^{\prime
})\equiv i\langle c_{q\sigma }^{\dagger }(t^{\prime })a_{k\sigma ^{\prime
}}(t)\rangle .$ By using the nonequilibrium Green function technique\cite
{zhu}, we can get the torque, $\tau ^{Rx^{\prime }}$, to the first order of
the Green function by 
\begin{equation}
\tau ^{Rx^{\prime }}=\pi \int d\varepsilon \lbrack f(\varepsilon
+eV)-f(\varepsilon )]Tr_{\sigma }[{\bf \Lambda (}\stackrel{\wedge }{\sigma }%
_{1}\cos \theta -\stackrel{\wedge }{\sigma }_{3}\sin \theta )],  \label{xt1}
\end{equation}
where $f(x)$ is the Fermi function, the matrix ${\bf \Lambda }$ is defined
as ${\bf \Lambda }={\bf T}^{\dagger }{\bf D}_{L}(\varepsilon +eV){\bf TRD}%
_{R}(\varepsilon ){\bf R}^{\dagger }=\left( 
\begin{array}{cc}
\Lambda _{1} & \Lambda _{2} \\ 
\Lambda _{3} & \Lambda _{4}
\end{array}
\right) ,\ $and ${\bf D}_{L(R)}=\left( 
\begin{array}{cc}
D_{L(R)\uparrow } & 0 \\ 
0 & D_{L(R)\downarrow }
\end{array}
\right) $ whose elements $D_{L(R)\uparrow (\downarrow )}(\varepsilon
)=D_{L(R)}(\varepsilon \pm M_{1(2)})$ are the density of states (DOS) of
electrons with spin up and down in the left (right) ferromagnet,
respectively. After some algebras, we have

\begin{equation}
\tau ^{Rx^{\prime }}=\frac{\pi }{2}\int d\varepsilon \lbrack f(\varepsilon
)-f(\varepsilon +eV)](D_{R\uparrow }+D_{R\downarrow })\Gamma
_{1}^{L}(P_{1}\sin \theta -P_{3}\cos \theta ),  \label{troque1}
\end{equation}
where $P_{1}=\frac{D_{L\uparrow }(T_{1}^{2}-T_{2}^{2})-D_{L\downarrow
}(T_{4}^{2}-T_{3}^{2})}{D_{L\uparrow }(T_{1}^{2}+T_{2}^{2})+D_{L\downarrow
}(T_{3}^{2}+T_{4}^{2})},$ $P_{3}=\frac{2(D_{L\uparrow
}T_{1}T_{2}+D_{L\downarrow }T_{3}T_{4})}{D_{L\uparrow
}(T_{1}^{2}+T_{2}^{2})+D_{L\downarrow }(T_{3}^{2}+T_{4}^{2})},$ and $\Gamma
_{1}^{L}=D_{L\uparrow }(T_{1}^{2}+T_{2}^{2})+D_{L\downarrow
}(T_{3}^{2}+T_{4}^{2}).$ It is interesting to note from the above equation
that the direction of the spin torque is closely related to whether the
applied bias is positive or negative, namely, it depends strongly on the
direction of the electrical current, in agreement of the previous
observation \cite{slonczewski,tsoi,sun,myers,katine}.

To compare the results with those reported in Refs. \cite
{slonczewski,waintal}, we can also consider the spin-torque per current,
i.e. $\tau ^{Rx^{\prime }}/I_{e}=\left\langle \frac{1}{G}\frac{\partial \tau
^{Rx\prime }}{\partial V}\right\rangle ,$ where $G$ is the tunneling
conductance [see Eq. (12) in Ref. \cite{zhu}] and $V$ is the applied bias.
Then we obtain 
\begin{equation}
\frac{\tau ^{Rx^{\prime }}}{I_{e}}=\frac{\hbar }{e}\frac{P_{1}\sin \theta
-P_{3}\cos \theta }{1+P_{2}(P_{1}\cos \theta +P_{3}\sin \theta )},
\label{torqandcurr}
\end{equation}
where $P_{2}=\frac{D_{R\uparrow }-D_{R\downarrow }}{D_{R\uparrow
}+D_{R\downarrow }}$ is the polarization of the right ferromagnet, and the
energy is taken at the Fermi level.

To gain deeper insight into the effect of the spin-flip scatterings on the
spin tranfer torque, we ought to invoke numerical calculations. Before
presenting the calculated results, we shall presume a parabolic dispersion
for band electrons based on which the DOS of conduction electrons are
calculated. The Fermi energy and the molecular field will be taken as $%
E_{f}=1.295$ eV and $\left| {\bf h}_{1}\right| =\left| {\bf h}_{2}\right|
=0.90$ eV, which are given in Ref. \cite{moodera1} for Fe. In addition, we
may for convenience introduce two parameters $\gamma _{1}=T_{2}/T_{1}$ and $%
\gamma _{2}=T_{3}/T_{1}$, and assume $T_{1}=T_{4}$.

Now let us first look at the case without the spin-flip scatterings. The
leading contribution to the torque $\tau ^{Rx^{\prime }}$ comes from the
first nonvanishing term. When the spin-flip scatterings are absent, i.e. $%
T_{2}=T_{3}=0,$ we find that $\frac{\partial \tau ^{Rx^{\prime }}}{\partial V%
}=\frac{e\pi }{2}(D_{R\uparrow }+D_{R\downarrow })(D_{L\uparrow
}+D_{L\downarrow })T_{1}^{2}\overline{P}_{1}\sin \theta $ with $\overline{P}%
_{1}=(D_{L\uparrow }-D_{L\downarrow })/(D_{L\uparrow }+D_{L\downarrow }),$
which is consistent with that obtained in Refs.\cite{slonczewski,s2}. This
result shows clearly that $\tau ^{Rx^{\prime }}$ vanishes when the relative
alignment of magnetizations of the two ferromagnets is parallel ($\theta =0$%
) or antiparallel ($\theta =\pi $). It is similar to that for a FM-NM-FM
trilayer system discussed in Ref.\cite{waintal}, although the transport
mechanisms are different. The present result can be easily understood
because the spin-polarized electrons along the $z$ or $-z$ axis cannot feel
the spin transfer torque owing to the property of $\widehat{s}_{1,2}\times (%
\widehat{s}_{1}\times \widehat{s}_{2})$. In Fig. 2, we show the $\theta $
dependence of the spin torque $\frac{\tau ^{Rx^{\prime }}}{I_{e}}$ in the
absence of spin-flip scatterings for different polarizations. One may see
that $\frac{\tau ^{Rx^{\prime }}}{I_{e}}$ is a nonmonotonic function of $%
\theta $, and shows minima and maxima at certain relative alignments. $\frac{%
\tau ^{Rx^{\prime }}}{I_{e}}$ versus $\theta $ is inversion-symmetrical to
the axis of $\theta =0$. It is evident that the larger the polarization, the
stronger the spin transfer torques. This observation is in good agreement
with the finding in Ref.\cite{slonczewski}, though the latter is obtained on
a basis of a quite different method.

The $\theta $-dependence of $\frac{\tau ^{Rx^{\prime }}}{I_{e}}$ for $\gamma
_{1}=\gamma _{2}$ is shown in Fig. 3 (a). It can be seen that the spin-flip
scatterings can lead to a nonvanishing spin torque at $\theta =0$ or $\pi ,$
i.e. $\frac{\partial \tau ^{Rx^{\prime }}}{\partial V}=\mp e\pi
(D_{R\uparrow }+D_{R\downarrow })T_{1}^{2}(\gamma _{1}D_{L\uparrow }+\gamma
_{2}D_{L\downarrow }),$ being different from the case when the spin-flip
scatterings are neglected. However, in the present case with $\gamma
_{1}=\gamma _{2}=0.05,$ at a particular angle, e.g. $\theta =14.04^{\circ }$%
, $\tau ^{Rx^{\prime }}$ becomes zero. This suggests that the spin-flip
scattering can lead to an angular shift to the spin torque. In other words,
the spin-flip scattering\ gives rise to an additional torque which we may
call the {\it spin-flip induced spin torque} henceforth. In addition, if $%
\gamma _{1}=\gamma _{2}=\gamma ,$ one may find that $\frac{\partial \tau
^{Rx^{\prime }}}{\partial V}$ is propotional to $\gamma $. When $\gamma
_{1}\neq \gamma _{2}$, which means that the spin-flip scatterings from the
spin up band to the spin down band are different from those from down to up, 
$\gamma _{1}$ and $\gamma _{2}$ give different effects on the $\theta $%
-dependence of spin torques as shown in Fig. 3 (b). It is observed that
there are angular shifts for different $\gamma _{1}$ and $\gamma _{2}$,
showing that the effects of $\gamma _{1}$ and $\gamma _{2}$ are various. For
instance, the curve for $\gamma _{1}=0.2$ and $\gamma _{2}=0.1$ moves to the
right-hand side in comparison to the curve of \ $\gamma _{1}=0.1$ and $%
\gamma _{2}=0.2$. It appears that a larger spin-flip scattering from spin
down to spin up has a larger effect on the spin torque.

The effect of the molecular fields of the ferromagnets on the spin torques
is also investigated. The $\theta $-dependences of $\tau ^{Rx^{\prime }}$
are shown in Fig. 4 for different parameter $\alpha =\left| {\bf h}%
_{R}\right| /\left| {\bf h}_{L}\right| $. One may find that $\tau
^{Rx^{\prime }}/I_{e}$ is also a nonmonotinoc function of $\theta $, and
shows peaks at certain values of $\theta $ for different $\alpha ^{\prime }$%
s. It is interesting to note that the two crossing points at $\theta
_{1}=37^{\circ }$ and $\theta _{2}=127^{\circ }$ under different molecular
fields are observed. When $\theta <$ $\theta _{1}$, a larger $\alpha $ leads
to a smaller magnitude of the spin-torque (which are negative); when $\theta
_{1}<\theta <\theta _{2}$, the larger $\alpha ,$ the smaller the
spin-torques; while $\theta >\theta _{2}$, a larger $\alpha $ leads to a
larger magnitude of the spin-torque, showing that $\left| {\bf h}_{R}\right| 
$ and $\left| {\bf h}_{L}\right| $ have different effects on the spin
torques. The two crossing points appeared in the curve $\tau ^{Rx^{\prime
}}/I_{e}$ versus $\theta $ for different molecular fields can be understood
in the following way. From Eq. (\ref{torqandcurr}), one can get a relation 
\begin{equation}
\frac{\tau ^{Rx^{\prime }}}{I_{e}}=\frac{\hbar }{e}\frac{\sqrt{%
P_{1}^{2}+P_{3}^{2}}\sin (\theta -\theta _{f})}{1+P_{2}\sqrt{%
P_{1}^{2}+P_{3}^{2}}\cos (\theta -\theta _{f})},  \label{anothert}
\end{equation}
where the angular shift $\theta _{f}$ is defined by $\tan \theta _{f}=\frac{%
P_{3}}{P_{1}}$\cite{zhu}. $\theta _{1}$ at which the first crossing point
appears is nothing but $\theta _{f}$ (i.e. $\theta _{1}=\theta
_{f}=37^{\circ }$). At $\theta _{1}=\theta _{f}$, $\frac{\tau ^{Rx^{\prime }}%
}{I_{e}}=0$, which is independent of the molecular field of the right
ferromagnet. The second crossing point appears at $\theta _{2}-\theta _{f}=%
\frac{\pi }{2}$ (i.e. $\theta _{2}=127^{\circ }$). At this angle $\theta _{2}
$, $\frac{\tau ^{Rx^{\prime }}}{I_{e}}=\frac{\hbar }{e}\sqrt{%
P_{1}^{2}+P_{3}^{2}}$, which depends only on the parameters of the left
ferromagnet but is independent of the molecular field of the right
ferromagnet. Therefore, at these two particular alignments, $\frac{\tau
^{Rx^{\prime }}}{I_{e}}$ gives the same value for different $\alpha ^{\prime
}$s, thereby leading to the two crossing points. When the spin-flip
scattering disappears, the two crossing points occur at $\theta =$ $0$ and $%
\frac{\pi }{2}$. So we may find that the spin-flip scattering may lead to an
additional spin torque on the magnetic moments of the right ferromagnet. The
present observation may be readily understood, because the spin-flip
tunneling of electrons from the left ferromagnet into the right ferromagnet
would give rise to a change of the DOS of conduction electrons as well as
the effective polarization factor in the right ferromagnet, leading to an
additional torque exerting on the magnetic moments, thereby the property of
an angular shift of the spin torque was observed.

In summary, we have investigated the current-induced spin torque in FM-I-FM
tunnel junctions with inclusion of spin-flip scattering by using the
nonequilibrium Green function method. In the absence of the spin-flip
scattering, our results are consistent with the previous results found in
Refs.\cite{slonczewski,s2,waintal}. When the spin-flip scattering, the
factor that could exist in realistic spin-based electronic devices, is
considered, we have found that an additional spin torque is induced. It is
found that the spin-flip scattering can enhance the maximum of the
current-induced spin transfer torque, giving rise to an angular shift. The
effects of the molecular fields of the left and right ferromagnets on the
spin torques are also studied. It can be observed that the spin-torques per
unit tunneling current exerting on the right ferromagnet are independent of
the molecular field of the right ferromagnetic lead at $\theta =\theta _{f}$
and $\theta -\theta _{f}=\frac{\pi }{2}$. The present study shows that the
spin-flip scatterings during the tunneling process in magnetic tunnel
junctions have indeed remarkable influences on the dynamics of the magnetic
moments. Since the spin-transfer torque induced by the applied current can
switch the magnetic domains between different orientations\cite{myers},
people can invoke this property to fabricate a current-controlled magnetic
memory element. As the previous studies on the spin torque usually ignore
the effect of the spin-flip scatterings which could not be avoided in
practice, our investigation might offer a supplement to the previous
studies, namely, when people design a device based on a mechanism of the
spin transfer torque, one should take the additional torque induced by the
spin-flip scatterings into account, which could complement with the
experiments.

\section*{Acknowledgments}

This work is supported in part by the National Science Foundation of China
(Grant No. 90103023, 10104015), the State Key Project for Fundamental
Research in China, and by the Chinese Academy of Sciences.

{\bf FIGURE CAPTIONS}

Fig. 1 A schematic illustration of the spin-transfer torque with spin-flip
scatterings in a $FIF$ tunnel junction.

Fig. 2 Spin transfer torque $\frac{\tau ^{Rx^{\prime }}}{I_{e}}$versus angle 
$\theta $ in the absence of the spin-flip scatterings (i.e.$\gamma
_{1}=\gamma _{2}=0$). Equal polarization factors of the magnets are assumed (%
$P_{1}=P_{2}$). Torque per unit current is measured in unit of $\hbar /e.$

Fig. 3 The $\theta $ dependence of $\frac{\tau ^{Rx^{\prime }}}{I_{e}}$ for
(a) $\gamma _{1}=\gamma _{2}$ and (b) $\gamma _{1}\neq \gamma _{2}$, where
the\ effective masses of the left and right ferromagnets are taken as unity,
the molecular fields are assumed to be $0.9$ eV, the Fermi energy is taken
as $1.295$ eV, and $T_{1}=T_{4}$ $=0.01$ eV. Torque per unit current is
measured in unit of $\hbar /e.$

Fig. 4 $\tau ^{Rx^{\prime }}/I_{e}$ versus angle $\theta $ for different
molecular fields. Here we take $\gamma _{1}=\gamma _{2}=0.15$, $\left| {\bf h%
}_{1}\right| =0.9$ eV, and the other parameters are taken the same as those
in Fig. 3. Torque per unit current is measured in unit of $\hbar /e.$

\end{document}